\documentclass[12pt,aps,preprint,nofootinbib]{revtex4}
\usepackage{graphicx}
\usepackage{bm}
\usepackage{amssymb,amsmath,amsfonts,amsbsy,epsfig}
\usepackage{color}
\usepackage{amsmath}
\usepackage{amssymb}
\usepackage{color}
\usepackage{cancel}
\def\beq{\begin{equation}}
\def\eeq{\end{equation}}                         
\def\bea{\begin{eqnarray}}
\def\eea{\end{eqnarray}} 
\def\to{\rightarrow}

\def\o(#1){\mathcal{O}(#1)}

\begin{document}

\preprint{IPMU10-0146} 
\title{Inverse Seesaw in Supersymmetry}
\author{Seong Chan Park~$^{a}$ and Kai Wang~$^{a,b}$}
\affiliation{$^{a}$~Institute for Physics and Mathematics of the Universe (IPMU), the University of Tokyo, Kashiwa, Chiba 277-8568, JAPAN\\
$^{b}$~Zhejiang Institute for Modern Physics (ZIMP), Zhejiang University, Hangzhou, Zhejiang 310027, CHINA}

\begin{abstract}
We study a mechanism where tiny neutrino masses arise only from radiative contribution in a supersymmetric model. In each generation, the tree-level light neutrino mass is rotated away by introducing a new singlet neutrino $s_{L}$ that forms a Dirac mass term with the right-handed neutrino $n_{R}$. With non-zero Majorana neutrino mass for the right-handed neutrinos $M_{R} \overline{n^{c}_{R}} n_{R}$, the lightest neutrino remains massless at tree level. 
Supersymmetry ensures that the Majorana neutrino masses  $M_{R} \overline{n^{c}_{R}} n_{R}$
and $M^{*}_{R} \overline{s^{c}_{L}} s_{L}$ are not generated simultaneously. 
There is no exact chiral symmetry to protect the neutrino mass.
Consequently, tiny neutrino masses then only arise from radiative contributions and 
the right-handed neutrino Majorana mass $M_{R}$ can be at $\cal O$(KeV).    
\end{abstract}
\maketitle

Enormous experimental evidences have shown that neutrinos have tiny masses of ${\mathcal O}(10^{-10})$~GeV. 
Being completely neutral under the unbroken gauge symmetries $SU(3)_{c}\times U(1)_{EM}$, neutrinos can
be Majorana fermions and the origin of neutrino mass may be different from the other SM fermions. 
In the minimal Higgs boson model, the Dirac neutrino masses can arise by introducing SM singlet fields $n^{i}_{R}$ but the fact that dimensionless Yukawa coupling $Y_{\nu}$ is of order ${\mathcal O}(10^{-12})$ is still puzzling.
The most elegant mechanism for neutrino mass generation is perhaps
the seesaw mechanism  \cite{GRSY,M,so10}.  Majorana masses of $n_{R}$ are introduced in addition to the Dirac terms as 
\beq
y_{\nu}\overline{\ell_{L}}{n_{R}} H + M_{R} \overline{n^{c}_{R}}n_{R}+h.c.~,
\eeq
where $\ell_{L}$ is the $SU(2)_{L}$ doublet $(1,2)_{-1}$, $y_{\nu} \langle H\rangle=M_{D}$ after electroweak breaking
and the light neutrinos get masses of order of $M^{T}_{D} M^{-1}_{R} M_{D}$. The tiny light neutrino masses are realized as a consequence of setting $M_{R}$ to ultra-high Grand Unification (GUT) scale \cite{GRSY} and the seesaw mechanism can
be naturally embedded into various GUT models. However, hierarchy problem arises since the heavy right-handed neutrinos contribute large logarithm corrections to the Higgs mass as $\Delta m^{2}_{h}\simeq y^{2}_{\nu} M^{2}_{R} \ln(q/M_{R})/4\pi^{2}$ \cite{Vissani:1997ys}.
It is then worth to investigating the possibility of explaining neutrino mass within weak scale. 
Radiative neutrino mass generation is one of the attempts in this approach. 
Neutrino masses can only be generated via radiative contribution in various models\cite{zee,babu}.
Some other models may contain right-handed neutrinos $n_{R}$ then the tree level mass
must be suppressed \cite{ma} so that the radiative contribution can dominate the neutrino mass. 
However, the existence of the radiative contributions in these models implies that 
the $U(3)_{\nu}$ chiral symmetry must be broken and the tree level mass cannot 
be set to zero by setting $y_{\nu}$. Therefore, in the models \cite{ma}, the Higgs 
vacuum expectation value is usually suppressed in the neutrino Yukawa interaction
to suppress the tree level contribution. In this paper, we discuss the possibility
of suppressing tree level mass without tuning the Yukawa coupling $y_{\nu}$ or
the Higgs vev. 

In the Dirac neutrino mass case,  if one introduces a new SM singlet $s_{L}$ to form another 
Dirac mass term with the right-handed neutrino $n_{R}$ as \cite{massless}
\beq
y_{\nu}\overline{\ell_{L}}n_{R} H +M_{S}\overline{s_{L}}n_{R}+h.c.~,\label{wyler}
\eeq
The model then contains one massless neutrino and one massive Dirac neutrino per generation. 
The ``inverse seesaw mechanism'' \cite{inverse} extended this model by introducing a Majorana mass term for
the $s_{L}$ states as 
\beq
y_{\nu}\overline{\ell_{L}}n_{R} H +M_{S}\overline{s_{L}}n_{R}+\epsilon \overline{s^{c}_{L}}s_{L}+h.c.~,
\eeq
The lightest neutrino mass then arises from a small $\epsilon$ as 
\beq
m_{\nu} \simeq \epsilon \frac{M^{2}_{D}}{M^{2}_{D}+M^{2}_{S}}
\eeq
while $\epsilon$ can be identified as soft breaking of $U(1)_{\rm Lep}$ lepton number symmetry. 

However, if one only introduces $M_{R} \overline{n^{c}_{R}}n_{R}$ to the Eq. \ref{wyler} as
in the seesaw mechanism,
\beq
y_{\nu}\overline{\ell_{L}}{n_{R}} H +M_{S}\overline{s_{L}}n_{R}+ M_{R} \overline{n^{c}_{R}}n_{R}+h.c.~,\label{lag}
\eeq
the lightest neutrino will remain massless at tree level and the original seesaw mechanism breaks down. To see this, we write down the neutrino mass matrix in the basis of
$(\nu_{L},  s_{L},n^{c}_{R})$
\begin{eqnarray}
{\cal M}= \left(
  \begin{array}{ c c  c}
     0 &  0 & M_{D}\\
     0 & 0 & M_{S}  \\
     M_{D} & M_{S} & M_{R}
  \end{array} \right)~.
  \label{matrix}
\end{eqnarray}
The mass eigenstates consist of one massless state and two massive states which are
mixture of Dirac and Majorana spinors
\bea
\nu &=& -\frac{M_{S}}{\sqrt{M^{2}_{D}+M^{2}_{S}}}\nu_{L}+ \frac{M_{D}}{\sqrt{M^{2}_{D}+M^{2}_{S}}}s_{L}\nonumber\\
N_{\pm} &=& \frac{1}{\sqrt{M^{2}_{\pm}+M^{2}_{D}+M^{2}_{S}}} ( M_{D}\nu_{L} + M_{S}s_{L}- M_{\pm}n^{c}_{R})
\eea
with mass eigenvalues as
\beq
m_{\nu}=0,~~M_{\pm}={1\over 2} \left(M_{R}\pm \sqrt{4M^{2}_{D}+M^{2}_{R}+4M^{2}_{S}}\right)~.
\eeq

We take one generation to illustrate the features of this model. For the symmetries that only act on neutral fermions, 
the lagrangian of free fields has accidental symmetries as $U(1)_{\nu}\otimes U(1)_{n}\otimes U(1)_{s}$. 
The existence of Dirac neutrino mass $M_{D}$ breaks $U(1)_{\nu}\otimes U(1)_{n}$ down to
$U(1)_{\nu+n}$ \footnote{If $M_{D}$ vanishes, the tree level neutrino mass is also massless. However, 
$y_{\nu}=0$ would restore the chiral symmetry and no neutrino mass would be generated radiatively as
long as there exists the exact chiral symmetry. On the other hand, the vacuum expectation value (VEV) 
$\langle H\rangle$ is usually associated with the up-type quark mass like $m_{t}$ and should not vanish.}
and $M_{S}$  also breaks $U(1)_{s}\otimes U(1)_{n}$ down to $U(1)_{s+n}$ \footnote{If $M_{S}$ vanishes, the model
simply becomes the original seesaw mechanism.}.
One can redefine the two $U(1)$ symmetries and identify one of $U(1)$s as  $U(1)_{\rm Lep}$ Lepton Number symmetry under which the fields transform as,
\beq
\nu_{L} \to e^{i\alpha} \nu_{L},~~s_{L} \to e^{i\alpha}s_{L},~~n^{c}_{R} \to e^{-i\alpha} n^{c}_{R}~.
\eeq
The second $U(1)$ can be identified as $U(1)_{\nu-s}$ under which $\nu_{L}$ and $s_{L}$ transform in the same way.
However, the $U(1)_{\nu-s}$ is only an approximate symmetry. By writing $U(1)_{\nu}$ instead of $U(1)_{\ell}$, we
did not assume the SM gauge symmetry while $\nu_{L}$ is charged under SM gauge group and $s_{L}$ is a completely SM singlet, $U(1)_{\nu-s}$ 
is not respected by the SM gauge interactions or interaction via the Higgs. 
Therefore, the only exact symmetry is $U(1)_{\rm Lep}$ which is later broken
by the $M_{R}$ explicitly \footnote{If the $U(1)_{\rm Lep}$ is exact symmetry, the lightest mass eigenstate will be exactly
massless to all orders.}.  One should also notice that at tree level, the Majorana masses of 
$\overline{\nu^{c}_{L}}\nu_{L}$, $\overline{s^{c}_{L} }s_{L}$
or $\overline{\nu^{c}_{L} }s_{L}$ all explicitly break the $U(1)_{\nu-s}$. 

In principle, once $M_{R} \overline{n^{c}_{R}}n_{R}$ term is generated, the term 
$M^{*}_{R} \overline{s^{c}_{L} }s_{L}$ will be generated automatically. 
To forbid the $M^{*}_{R}\overline{s^{c}_{L}}s_{L}$ term,
a natural extension is to embed the model into supersymmetric theory.
The holomorphic feature of superpotential naturally split
the two terms so that they will not be generated simultaneously. Notice 
supersymmetry does not forbid the $\overline{s^{c}_{L}}s_{L}$ term.
We want to emphasize that this model is only technically natural as
a explicit breaking of $U(1)_{\rm Lep}$ $M_{R} \overline{n^{c}_{R}}n_{R}$
won't automatically generate the Majorana mass of $s_{L}$ field.

\section{Model}
The superpotential of the model contains
\beq
W\ni y_{\nu}\ell N^{c} H_{u}+y_{e}\ell E^{c} H_{d} +\mu H_{u} H_{d}+M_{S} S N^{c} + M_{R} N^{c}N^{c}. 
\eeq
where $N^{c}$, $E^{c}$, $S$ are the chiral superfields. We assume there exists only one explicit breaking
of $U(1)_{\rm Lep}$ as $M_{R} N^{c} N^{c}$ and no $SS$ breaking.
In Table \ref{charge}, the $R$-charge and the lepton number  $U(1)_{\rm Lep}$ charges
of the fields in the model have been given. Without losing generality, we write the
$R$-charge in $SU(5)$ compatible language. 
\begin{table}[h]
\begin{tabular*}{0.75\textwidth}{@{\extracolsep{\fill}}c||c c c c | c c | c}
\hline
Field & $\ell$ & $E^{c}$ & $N^{c}$ & $S$ & $H_{u}$ & $H_{d}$ & $\theta$ \\
\hline
\hline
$R$-charge &  1/5 & 3/5 & 1 & 1 & 4/5 & 6/5 & 1\\  
$U(1)_{\rm lep}$   & 1 & -1 & -1 & 1 & 0 & 0 & 0\\
\hline
\end{tabular*}
\caption{Charge assignment of leptons and Higgses in the Model}

\label{charge}
\end{table}%

To ensure non-zero eigenvalues of Eq. \ref{matrix}, at least one of the Majorana mass terms 
\beq
\overline{\nu^{c}_{L}} s_{L}  + \overline{s^{c}_{L}} s_{L}  + \overline{\nu^{c}_{L}}\nu_{L}~
\eeq
should be at present. We summarize the properties of these terms in Table \ref{terms}, 
\begin{table}[h]
\begin{center}
\begin{tabular*}{0.75\textwidth}{@{\extracolsep{\fill}}c c c c}
\hline
 $W_{\rm eff}$ & $\cal L$ & $R$-charge of $\cal L$ & $U(1)_{l}$ charge\\
\hline
$N^{c} N^{c}$ & $\overline{n^{c}_{R}} n_{R}$ & $1+1-2\theta=0$ & -2 \\
\hline
$\ell S H_{u}$ & $\overline{\nu^{c}_{L}} s_{L}$ & ${1\over 5}+1 +{4\over 5} -2\theta=0$ & 2\\
$\ell \ell H_{u} H_{u}$ & $\overline{\nu^{c}_{L}}\nu_{L}$ & ${1\over 5}+{1\over 5}+{4\over 5}+{4\over 5}-2\theta=0$ & 2\\
$SS$ & $\overline{s^{c}_{L}}s_{L}$ & $1+1-2\theta=0$ & 2\\
\hline
\end{tabular*}
\end{center}
\caption{Properties of Lepton number violation terms in the model. $\cal L$ stands for the fermion mass terms in the lagrangian. }
\label{terms}
\end{table}%
If the right-handed Majorana mass $\overline{n^{c}_{R}}n_{R}$ carries $U(1)_{\rm Lep}$ Lepton number charge as $-2$,
the terms $\overline{\nu^{c}_{L}}\nu_{L}$, $\overline{s^{c}_{L}}s_{L}$ or $\overline{\nu^{c}_{L}}s_{L}$ 
all carry lepton number $+2$. None of the Majorana neutrino mass terms 
breaks $R$-symmetry. Therefore, the corresponding soft SUSY breaking terms 
\beq
{\cal L}_{\rm soft} \ni B_{R}\tilde{n}_{R} \tilde{n}_{R}+ B_{\nu}  \tilde{\nu}_{L}\tilde{\nu}_{L}+B_{S}  \tilde{s}_{L}\tilde{s}_{L} + A_{s} \tilde{\ell}_{L} \tilde{s}_{L} H_{u} \label{soft}
\eeq
all break $U(1)_{\rm Lep}$ as well as the $R$-symmetry and the fermion masses due to Eq. \ref{soft} then 
require loops involving gaugino mass insertion.

A very similar philosophy has been employed in the uplifted MSSM \cite{uplift}.
The holomorphic feature of superpotential and the anomaly 
cancellation conditions require the supersymmetric SM to be a 
Two-Higgs-doublet-model (2HDM) with $\langle H_{u}\rangle$ and $\langle H_{d}\rangle$
responsible for generation of $m_{u}$ and $m_{d} (\text{or}~~m_{e})$ respectively.
In MSSM, contributions to $m_{e}$ or $m_{d}$ from the $\langle H_{u}\rangle$ are
then similar to our model. In that case, all the chiral symmetries associated with the SM fermions
are broken by the Yukawa couplings. When Peccei-Quinn (PQ) symmetry and 
$R$-symmetry have been broken, effective operators 
\beq
y^{\prime}_{d} Q d^{c} H^{\dagger}_{u} + y^{\prime}_{e} \ell_{L} e^{c} H^{\dagger}_{u}
\eeq
can then be radiatively generated.  

In the presence of the new singlet $s_{L}$ which is charged under $U(1)_{\rm lep}$, the gravitational anomaly (Tr[$U(1)_{B-L}$]) and 
cubic anomaly $[U(1)_{B-L}]^{3}$ become non-vanishing and $U(1)_{B-L}$ symmetry is then
anomalous \cite{rparity}. In minimal model, there is no additional gauge interaction besides the SM gauge symmetries. 

\section{Neutrino Masses}

To realize neutrino masses, one should look at the corrections to the zero entries
in the Eq .\ref{matrix}. Even though there is no symmetry to protect these terms 
from being generated, the magnitude of these terms depends on mediation. 
As being argued, with the second SM singlet $s_{L}$ which is 
also charged under $U(1)_{\rm Lep}$, the $U(1)_{B-L}$ can no longer be gauge symmetry
in the minimal model and the model cannot be embedded into minimal $E_{6}$, 
there is no other $U(1)'$ gauge interactions.  The $n_{R}$ and $s_{L}$ fields then 
become completely gauge singlet. The gauge singlet $s_{L}$ only couple through gravity interaction. 
Consequently, the terms involving $s_{L}$ field as $\overline{\nu^{c}_{L}}s_{L}$ or $\overline{s^{c}_{L}}s_{L}$ 
will then only arise from gravity mediation with $1/M_\text{Pl}$ suppression. 
The leading correction in Eq. \ref{matrix} is then $\overline{\nu^{c}_{L}}\nu_{L}$  type which
originates from $\ell_{L}\ell_{L} H_{u} H_{u}$.

For exactly the same reason,  the lepton number violation terms in the soft SUSY breaking 
lagrangian Eq. \ref{soft} can not be generated via gauge mediation but only from gravity
interaction. The bilinear lepton number violation $B$-terms are then of gravitino mass ${\cal O}(m_{3/2})$. 
The $R$-breaking contributions to neutrino masses are realized at the order of $m^{2}_{3/2}/M_{\lambda}$
where the $M_{\lambda}$ is the wino or bino masses. \footnote{
In gauge mediated SUSY breaking (GMSB) models, it would be interesting
if the $R$-breaking contribution were the leading since the tiny neutrino mass would then
be identified as a consequence of gravitational interaction with $1/M_{\rm Pl}$ 
suppression. However, $R$-breaking contribution is not the leading contribution here.}

\begin{figure}[h]
\includegraphics[scale=1.0,width=5cm]{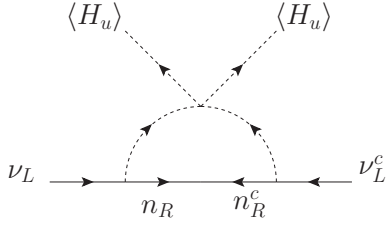}
\caption{One loop contribution of neutrino mass in non-SUSY. }
\label{diag1}
\end{figure}

To generate neutrino masses radiatively in this model, 
the chiral symmetries $U(1)_{\nu}\times U(1)_{n}\times U(1)_{s}$ as well as their
remanent $U(1)_{\rm Lep}$ must be broken. As being argued, the $U(1)_{\nu-s}$ is only an approximate 
symmetry since $\nu_{L}$ and $s_{L}$ can be easily distinguished through SM gauge interaction 
or interacting with Higgs, for instance the one-loop contribution as in Fig. \ref{diag1}. The processes 
involving weak gauge boson can be realized at even higher orders.

Figure \ref{diag1} shows that the contribution is proportional to the Yukawa coupling $y_{\nu}$ squared
as $y_{\nu}$ appears in two vertices. The loop induced mass vanishes in 
the limit of vanishing $y_{\nu}$ which restores $U(1)_{\nu}\times U(1)_{n}$ symmetries. 
$M_{R}$ dependence has two pieces. There is one mass insertion as $M_{R}$ and
the fermion propagators of $n$ fields also contain $M_{R}$. 
For simplicity, we want to discuss only the qualitative feature here. 
The neutrino mass due to Fig. \ref{diag1} is proportional to 
\beq
M_{\nu} \propto {1\over 16\pi^{2}} \lambda y^{2}_{\nu} \langle H_{u}\rangle^{2} M_{R} \sum_{\phi_{i}} \frac{R_{i}}{M^2_{\phi_{i}}-M^{2}_{R}} \ln\left(\frac{M^{2}_{\phi_{i}}}{M^{2}_{R}}\right)~.
\eeq
In evaluating the loop contribution in Fig. \ref{diag1}, one should compute the diagrams in the mass eigenstates. 
In MSSM, the $H_{u}$ is actually a mixture state of several neutral scalars $h,~H,~A$ and they can all run into the loops.
We use $R_{i}$ to denote the mixing factor for each scalar $\phi_{i}$. $\lambda$ stands for the Higgs 
quartic coupling which is $(g^{2}_{1}+g^{2}_{2})/8$ in MSSM. 
The interesting feature of this radiatively generated mass is that $M_{R}$ appears in both numerators and denominators as
\beq
\frac{M_{R}}{M^{2}_{\phi_{i}}-M^{2}_{R}}~.\label{nonsusy}
\eeq
With the scalar masses $M_{\phi_{i}}\sim M_{\rm EW}$ of ${\cal O}(10^{2})$~GeV, to obtain the tiny neutrino mass, one can take two different limit of $M_{R}$:                                                                                                                                                                                                                                                                                                                                                                                                                                                                                                                                                                                                                                                                                                                                                                                                                                                                                                                                                                                                                                                                                                                                                                                                                                                                                                                                                                                                                                                                                                                                                                                                                                                                                                                                                                                                                                                                                                                                                                    
\begin{itemize}
\item $M_{R}\gg M_{\phi_{i}}$ 
where correction then reduces to 
\beq
M_{\nu}\propto y^{2}_{\nu} \langle H_{u}\rangle^{2}/ M_{R}
\eeq
as in the conventional seesaw mechanism with additional loop suppression.
\item $M_{R}\ll M_{\phi_{i}}$
where the 
\beq
M_{\nu} \propto M_{R}~.
\eeq
Since there is additional loop suppression, it requires $M_{R}\sim {\cal O}$(KeV). 
\end{itemize}

\begin{figure}[h]
\vspace{0.5cm}
\includegraphics[scale=1.0,width=5cm]{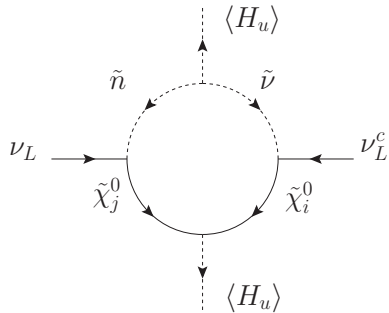}
\caption{One loop contribution of neutrino mass in SUSY. }
\label{diag2}
\end{figure}

Without breaking the $R$-symmetry, 
this supersymmetric model has another Lepton number violation vertex which is proportional to $M_{R}$.
Given the superpotential of the model as
\beq
W= \ell N^{c} H_{u} + M_{R} N^{c} N^{c} + M_{S} S N^{c}~.
\eeq
In the scalar potential, 
\begin{eqnarray}
V&=& \left| {\partial W \over \partial N^{c}}\right|^{2}= |(\tilde{\ell} H_{u}+M_{S}\tilde{s}+M_{R}\tilde{n})|^{2}\nonumber\\
&\ni& M^{*}_{R} \tilde{n}^{*}\tilde{\ell}H_{u}+M^{*}_{R} M_{S}\tilde{n}^{*}\tilde{s}~.
\end{eqnarray}
Both $M^{*}_{R} \tilde{n}^{*}\tilde{\ell}H_{u}$ and $M^{*}_{R} M_{S}\tilde{n}^{*}\tilde{s}$ are proportional to 
$M_{R}$ and violate $U(1)_{\rm Lep}$. However, since the sfermion $\tilde{s}$ do not participate in the gauge interactions
as we have argued, the contribution is then suppressed. 
The contribution to neutrino mass due to the vertex $M^{*}_{R} \tilde{n}^{*}\tilde{\ell}H_{u}$ is shown in Fig. \ref{diag2}.

Again, the Yukawa coupling $y_{\nu}$ appears in the vertex of Higgsino/right-handed sneutrino and
the vanishing $y_{\nu}$ leads to the vanishing mass. Figure \ref{diag2} contributions also involve the 
right-handed sneutrinos in the loops and
the right-handed sneutrino mass is \footnote{Since the right-handed neutrino is completely gauge singlet, the right-handed sneutrino mass will not receive 
a $M_{\rm SUSY}$ level contribution if the SUSY breaking is not gravity mediated.}
\beq
m^{2}_{\tilde{n}} \simeq M^{2}_{R}~.
\eeq
$M_{R}$ will again appear in both vertices and propagators
and the behavior is very similar to the previous case in Eq. \ref{nonsusy}.

In this paper, we want to take the second limit as $M_{R} \ll M_{\rm SUSY}\sim M_{\rm EW}$
for various reasons. First of all, it is to ensure there is no large correction in K\"{a}lher potential to Majorana mass terms involving $s_{L}$
fields without making any additional assumption. 
Secondly, with weak scale $M_{S}$ and $M_{D}$, a tiny $M_{R}$ not only explains the light neutrino mass but also
predicts two nearly degenerated weak scale pseudo-Dirac neutrinos which can in principle be produced at the CERN Large Hadron Collider (LHC).
In the end, as being argued in ``inverse seesaw mechanism''\cite{inverse}, a tiny $M_{R}$ can be
identified as a soft breaking of $U(1)_{\rm Lep}$.

\section{Conclusions}

In this paper, we study the scenario where tiny neutrino mass arises as radiatively corrections. 
To suppress the tree level mass without restoring the chiral symmetry or tuning the VEV,
we introduce one singlet field in addition to the right-handed neutrino and rotate away 
the tree level  neutrino mass. We employ the same philosophy as in the inverse seesaw model that
there exists a tiny scale around KeV due to the soft breaking of $U(1)_{\rm Lep}$ Lepton number symmetry. 
Unfortunately, the spectrum predicted in this model is almost identical to the inverse seesaw mechanism
and it will be difficult to distinguish this model from the inverse seesaw model. 

\section*{Acknowledgement}
We would like to thank T. Yanagida for discussions and comments on this project 
and early collaboration on a previous related project \cite{Park:2009cm}. KW would also 
like to thank Jason Evans, Ming-xing Luo and Serguey Petcov for useful discussion.
The work is supported by the World Premier International Research Center Initiative (WPI Initiative), MEXT, Japan. 
SCP and KW are also supported in part by the JSPS Grant-in-Aid for scientific research (Young Scientists)
(B) 21740172 and (B) 22740143 respectively.

\end{document}